\documentclass[twoside,twocolumn,9pt]{article}
\usepackage{extsizes}
\usepackage[super,sort&compress,comma]{natbib} 
\usepackage[version=3]{mhchem}
\usepackage[left=1.5cm, right=1.5cm, top=1.785cm, bottom=2.0cm]{geometry}
\usepackage{balance}
\usepackage{mathptmx}
\usepackage{sectsty}
\usepackage{graphicx} 
\usepackage{lastpage}
\usepackage[format=plain,justification=justified,singlelinecheck=false,font={stretch=1.125,small,sf},labelfont=bf,labelsep=space]{caption}
\usepackage{float}
\usepackage{fancyhdr}
\usepackage{fnpos}
\usepackage[english]{babel}
\addto{\captionsenglish}{%
	
}
\usepackage{array}
\usepackage{droidsans}
\usepackage{charter}
\usepackage[T1]{fontenc}
\usepackage[usenames,dvipsnames]{xcolor}
\usepackage{setspace}
\usepackage[compact]{titlesec}
\usepackage{hyperref}
\usepackage{epstopdf}
\definecolor{cream}{RGB}{222,217,201}
\begin{document}	
	\pagestyle{fancy}
	\thispagestyle{plain}
	\fancypagestyle{plain}{
		\renewcommand{\headrulewidth}{0pt}
	}
	
	\makeFNbottom
	\makeatletter
	\renewcommand\LARGE{\@setfontsize\LARGE{15pt}{17}}
	\renewcommand\Large{\@setfontsize\Large{12pt}{14}}
	\renewcommand\large{\@setfontsize\large{10pt}{12}}
	\renewcommand\footnotesize{\@setfontsize\footnotesize{7pt}{10}}
	\makeatother
	
	\renewcommand{\thefootnote}{\fnsymbol{footnote}}
	\renewcommand\footnoterule{\vspace*{1pt}%
		\color{cream}\hrule width 3.5in height 0.4pt \color{black}\vspace*{5pt}} 
	\setcounter{secnumdepth}{5}
	
	\makeatletter 
	\renewcommand\@biblabel[1]{#1}            
	\renewcommand\@makefntext[1]%
	{\noindent\makebox[0pt][r]{\@thefnmark\,}#1}
	\makeatother 
	\renewcommand{\figurename}{\small{Fig.}~}
	\sectionfont{\sffamily\Large}
	\subsectionfont{\normalsize}
	\subsubsectionfont{\bf}
	\setstretch{1.125} 
	\setlength{\skip\footins}{0.8cm}
	\setlength{\footnotesep}{0.25cm}
	\setlength{\jot}{10pt}
	\titlespacing*{\section}{0pt}{4pt}{4pt}
	\titlespacing*{\subsection}{0pt}{15pt}{1pt}
	\fancyhead{}
	\renewcommand{\headrulewidth}{0pt} 
	\renewcommand{\footrulewidth}{0pt}
	\setlength{\arrayrulewidth}{1pt}
	\setlength{\columnsep}{6.5mm}
	\setlength\bibsep{1pt}
	
	\makeatletter 
	\newlength{\figrulesep} 
	\setlength{\figrulesep}{0.5\textfloatsep} 
	
	\newcommand{\topfigrule}{\vspace*{-1pt}%
		\noindent{\color{cream}\rule[-\figrulesep]{\columnwidth}{1.5pt}} }
	
	\newcommand{\botfigrule}{\vspace*{-2pt}%
		\noindent{\color{cream}\rule[\figrulesep]{\columnwidth}{1.5pt}} }
	
	\newcommand{\dblfigrule}{\vspace*{-1pt}%
		\noindent{\color{cream}\rule[-\figrulesep]{\textwidth}{1.5pt}} }
	
	\makeatother
	
	%
	\twocolumn[
	\begin{@twocolumnfalse}
		{
			
		}\par
		\vspace{1em}
		\sffamily
		\begin{tabular}{m{1.5cm} p{14.5cm} }
			
			&\noindent\LARGE{\textbf{Shear jamming and fragility in fractal suspensions under confinement}} \\
			\vspace{0.3cm} & \vspace{0.3cm} \\
			
			& \noindent\large{Sarika C. K.,\textit{$^{a}$} Sayantan Majumdar,\textit{$^{a, *}$} and A. K. Sood \textit{$^{b}$}} \\
			\vspace{0.3cm} & \vspace{0.3cm} \\
			
			& \noindent \textit{$^{a}$~Soft Condensed Matter Group, Raman Research Institute, Bengaluru 560080, India.}\\
			& \noindent \textit{$^{b}$~Department of Physics, Indian Institute of Science, Bengaluru 560012, India.}\\
			& \noindent (Dated \today)\\
			\vspace{0.3cm} & \vspace{0.3cm} \\
			
			& \noindent\normalsize{Under applied stress, the viscosity of many dense particulate suspensions increases drastically, a response known as discontinuous shear-thickening (DST). In some cases, the applied stress can even transform the suspension into a solid-like shear jammed state. Although shear jamming (SJ) has been probed for dense suspensions with particles having well-defined shapes, such a phenomenon for fractal objects has not been explored. Here, using rheology and in situ optical imaging, we study the flow behaviour of ultra-dilute fractal suspensions of multi-walled carbon nanotubes (MWCNT) under confinement. We show a direct transition from flowing to SJ state without a precursory DST in fractal suspensions at an onset volume fraction, $\phi \sim$ 0.5\%, significantly lower than that of conventional dense suspensions ($\phi \sim$ 55\%). The ultra-low concentration enables us to demonstrate the fragility and associated contact dynamics of the SJ state, which remain experimentally unexplored in suspensions. Furthermore, using a generalized Wyart-Cates model, we propose a generic phase diagram for fractal suspensions that captures the possibility of SJ without prior DST over a wide range of shear stress and volume fractions.} \\
			
		\end{tabular}
		
	\end{@twocolumnfalse} \vspace{1.0cm}
	
	]
	
	\footnotetext{\textit{$^{*}$~smajumdar@rri.res.in}}
	\renewcommand*\rmdefault{bch}\normalfont\upshape
	\rmfamily
	\section*{}
	\vspace{-1cm}
	
	\section{Introduction}
	Dense particulate suspensions formed by dispersing inorganic or polymeric particles in Newtonian fluids show an array of striking non-Newtonian flow properties at high volume fractions. Among these, shear-induced increase in viscosity, commonly known as shear thickening, has attracted significant recent interest from experimental as well as theoretical points of view. During shear thickening, the increase in viscosity under increasing applied stress can be mild or abrupt, giving rise to continuous shear thickening (CST) or discontinuous shear thickening (DST) respectively \cite{barnes1989shear, hoffman1972discontinuous, mewis2012colloidal, wagner2009shear, brown2014shear}. Many of the dense suspensions showing DST also exhibit shear jamming (SJ) at high enough applied stress ($\tau$) and volume fraction ($\phi$) where the system develops finite yield stress like a solid. Recent studies point out that beyond a critical onset stress for contact formation, system-spanning contact networks similar to those observed for dry granular materials can explain the phenomenon of SJ in dense suspensions \cite{brown2014shear, bi2011jamming, wyart2014discontinuous, seto2013discontinuous}. Importantly, the SJ state is reversible: once the applied stress is removed, the system goes back to the initial fluid-like state in a short span of time \cite{barik2022origin}. Such reversible tuning of viscosity implies many potential applications, particularly in the field of designing smart shock-absorbing materials \cite{lee2003ballistic, majumdar2013optimal}. 
	
	In general, SJ can be observed only over a limited range of volume fractions: $\phi_\mu<\phi<\phi_0$, where $\phi_\mu$ and $\phi_0$ are the critical jamming packing fractions with and without friction respectively. For suspensions in which the friction and other stress-induced short-range interactions between the particles are negligible, $\phi_\mu$ remains close to $\phi_0$ and the range of $\phi$ over which shear jamming is observed diminishes. With greater friction, $\phi_\mu$ reduces and moves away from $\phi_0$ enabling SJ in suspensions over a considerable range of concentrations. Recent numerical studies \cite{guy2018constraint, singh2020shear} show that introducing constraints on the sliding and rolling motion of the particles by increasing the inter-particle friction and adhesion can lower $\phi_\mu$. Non-spherical particle shapes can also constrain the rolling motion and consequently reduce $\phi_\mu$ \cite{rathee2020role, james2019controlling}. Furthermore, studies show that tuning hydrogen bonding and solvent molecular weight enhances the particle contact interaction eliciting shear jamming in conventional suspensions \cite{james2019controlling, james2018interparticle, van2021role, peters2016direct}. In colloidal suspensions, it is demonstrated that the surface roughness of particles enhances DST due to the interlocking of surface asperities \cite{hsu2018roughness, hsu2021exploring}. A recent study shows that the critical volume fractions for both CST and DST can be lowered by tuning the surface chemistry and surface roughness of rigid particles in suspensions \cite{bourrianne2022tuning}. The key aspect emerging from these observations is the governing role of the contact network structure and the frictional contact strength between particles in the shear jamming of suspensions. Hence, a fractal suspension of stiff particles is a model system to probe shear jamming due to the strong frictional interaction between the particles. Confinement can facilitate the formation of the characteristic system-spanning network for shear jamming. With significantly high effective surface roughness and occupation volume, fractal suspensions are expected to have much lower $\phi_\mu$ compared to the non-fractal suspensions mentioned above.

	In this paper, we study the rheological behaviour of very dilute suspensions of multiwalled carbon nanotubes (MWCNT) dispersed in a Newtonian solvent, N-methyl-2-pyrrolidone (NMP), which form fractal aggregates (will be referred as flocs) due to cohesive interparticle interactions. Previously, the colossal increase in viscosity in this system by almost four orders of magnitude was thought to be due to DST \cite{majumdar2011discontinuous}. The present study establishes that this divergence in viscosity has all the signatures of shear jamming transition. For dense suspensions of frictional, non-fractal compact grains, as $\phi$ increases from a very low value, the system progressively passes through Newtonian/shear thinning (flowing), CST, DST, SJ and isotropically jammed (IJ) phases \cite{peters2016direct, fall2015macroscopic}. Equivalently, for dry granular systems the sequence of phases is: unjammed, fragile, SJ and IJ \cite{otsuki2020shear}. Interestingly, we observe a direct transition from flowing to SJ state with increasing $\phi$, completely bypassing any intermediate CST/DST phases. Moreover, the ultra-low concentration range ($\phi \geq 0.5\%$) for observing the SJ in the system allows us to experimentally probe, for the first time, the fragility and the associated structural reorganization in shear jammed suspensions. 
	
	\section{Results and Discussion}
	\begin{figure}
		\centering
		\includegraphics[width=1\columnwidth]{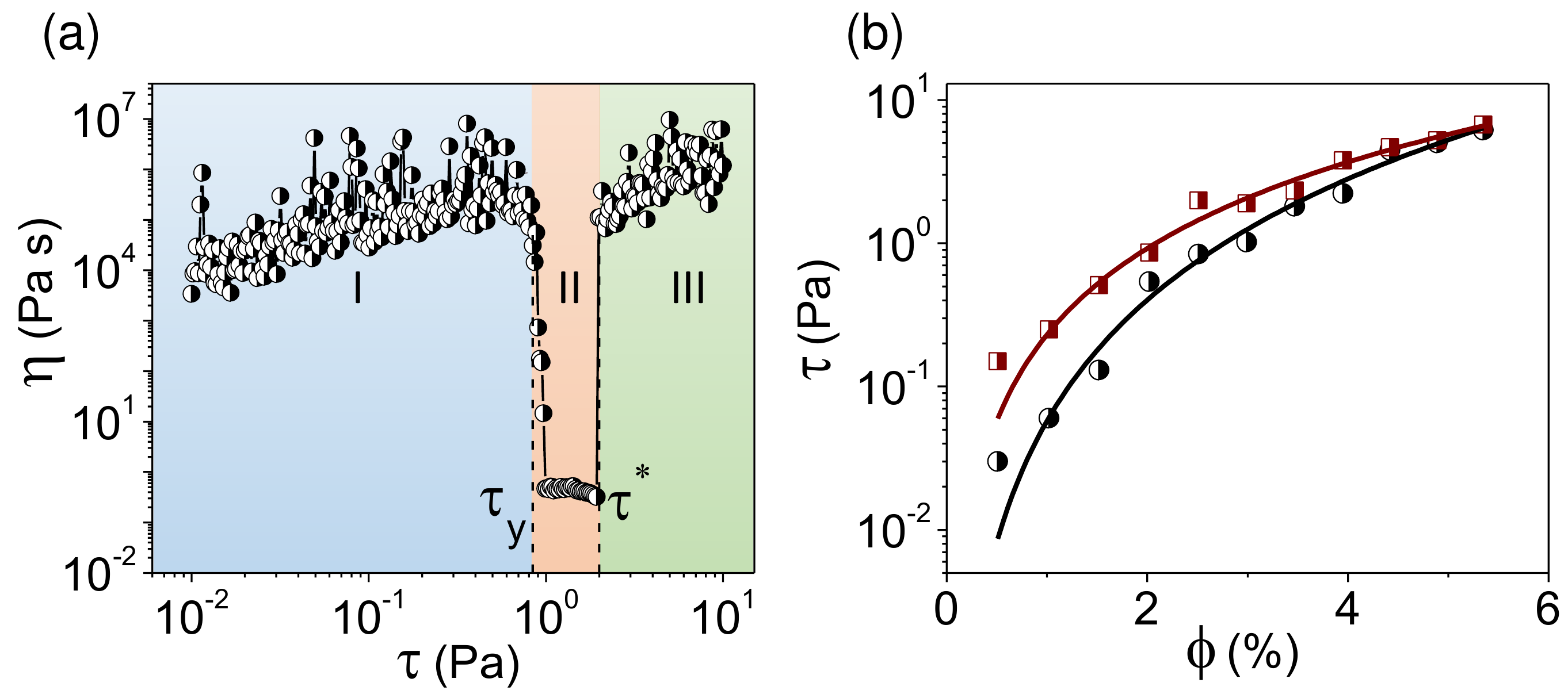}
		\caption{a) Flow curve of MWCNT suspension ($\phi$ = 2.5\%), plotted as viscosity versus applied shear stress in log-log scale. The duration of each measuring point of shear stress is 100 s. Three different states, CJ-state, flowing-state and SJ-state are marked as I, II and III respectively, in different colors. b) Variation of yield stress (circle) and shear jamming stress (square) with respect to $\phi$, along with power law fit (solid lines).}
		\label{FlowCurve}
	\end{figure}
	We measure the flow curve, viscosity ($\eta$) versus shear stress ($\tau$), of flocculated fractal suspensions of MWCNT under controlled shear stress using parallel plate geometry of the rheometer. The detailed fractal structure analysis of MWCNT flocs is described in the Supplementary Information (S.I.) and in Fig. S1 where we obtain 2-D fractal dimension of the dried clusters $\sim$ 1.5 - 1.6 indicating that the flocs have a much larger effective volume compared to the actual volume of the constituent MWCNTs. We consider a range of volume fractions varying from 0.26 to 5.35\% to measure the flow curves. We show a typical flow curve for $\phi$ = 2.5\% in Fig. \ref{FlowCurve}a. The complete set of flow curves is shown in Fig. S2. The flow curve exhibits three distinct rheological signatures over the shear stress range of 0.01 - 10 Pa. The suspension is initially in a jammed (unyielded) state (Region I) due to cohesive interaction between the flocs until the yield stress ($\tau_y$) is reached. We term this region as the cohesive jammed state (CJ state). Beyond $\tau_y$, the suspension begins to flow and shear thins upon a  further increase in $\tau$ (Region II). The flowing-state continues until the viscosity shoots up instantaneously at  higher shear stress, $\tau^*$ (Region III). The flow curves plotted as shear stress versus shear rate in Fig. S3a and b show that in both the regions I and III, the shear rate drops down to zero, reaching the resolution limit of the rheometer and fluctuates around zero with positive and negative values. This indicates that regions I and III are solid-like jammed states as the average shear rate remains negligible in these regions. Typically, for systems showing DST and SJ, the stress dependence of viscosity in the shear-thickening regime is given by a power-law behaviour: $\eta \sim \tau^{\delta}$, with  $\delta = 1$ marking the onset of DST. At higher stress values, $\delta$ becomes larger than 1 indicating unstable flows in the system due to shear-induced local jammed regions. In this regime, when the flow curve is represented by shear rate ($\dot{\gamma}$) versus stress ($\tau$), the so-called `S-shaped' curves \cite{wyart2014discontinuous} are obtained showing a negative slope ($\frac{d\, \dot{\gamma}}{d\, \tau} < 0$) just beyond the onset stress for shear thickening. At higher stress values the system becomes solid-like with a vanishing $\dot{\gamma}$. As $\dot{\gamma} \rightarrow 0$ at a finite $\tau$, the viscosity diverges at this point. Hence, region III of the flow curve manifests clear signatures of SJ transition for $\tau \geq \tau^*$. For solid-like CJ and SJ states, $\eta \rightarrow \infty$. However, the observed finite values of viscosity for the CJ and SJ states (Fig. \ref{FlowCurve}a) is an artifact originating from the instrumental limitation in measuring very small shear rates. Thus, rotational rheometry is not sufficient to characterize the CJ and SJ states and one requires oscillatory rheology for such quantification. Oscillatory rheology measurements reveal that both the CJ and SJ states are viscoelastic solid-like in nature (storage modulus $G'$ much higher than the loss modulus $G''$), with the magnitude of $G'$ for the SJ state much higher compared to that for the CJ state \cite{majumdar2011discontinuous}. Nonetheless, stress-controlled rotational rheometry reliably quantifies the flowing state, as well as, the yielding and SJ transitions that remain the main focus of the present study.
	
	\begin{figure}
		\centering
		\includegraphics[width=1\columnwidth]{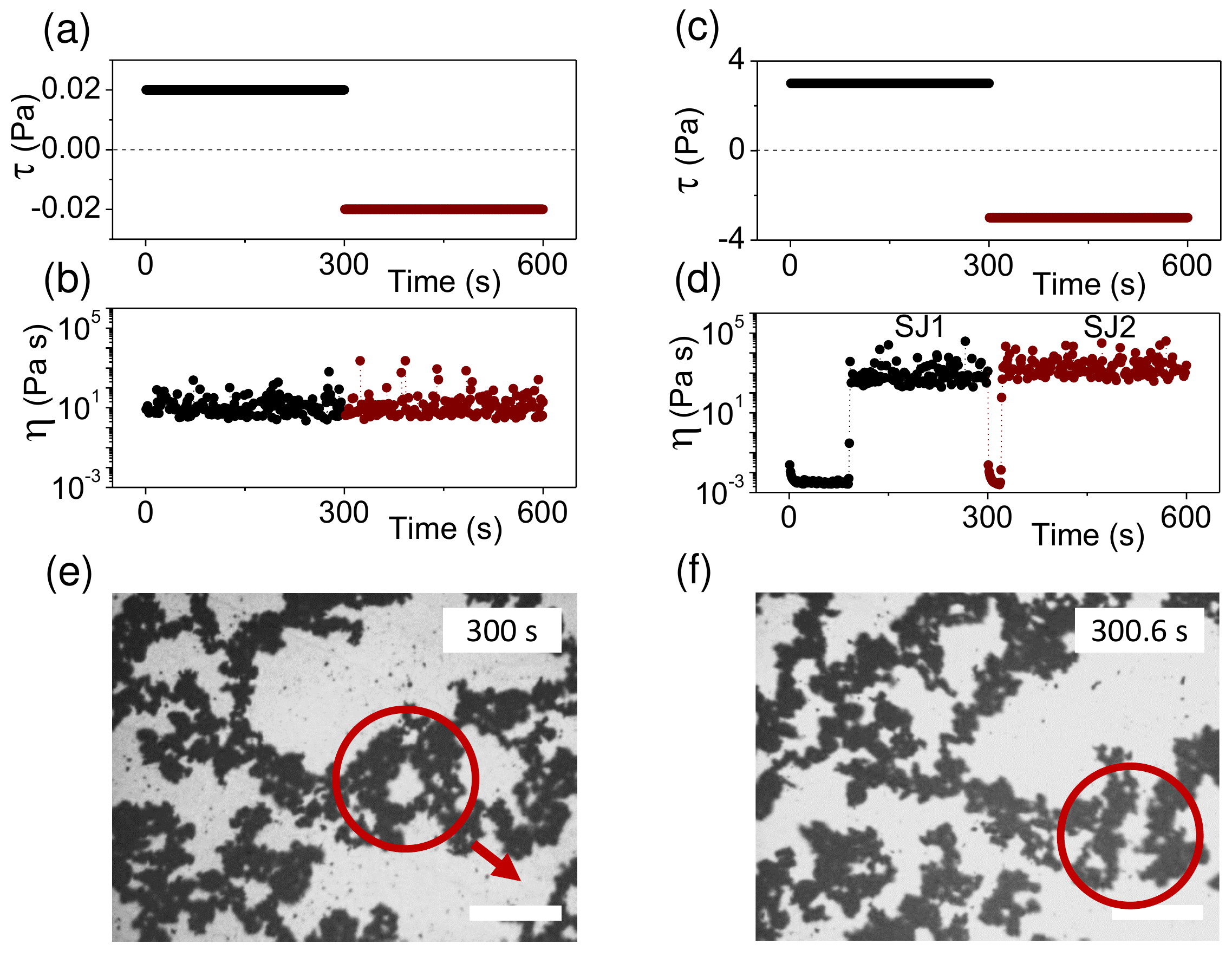}
		\caption{Stress reversal in CJ state (a, b) and SJ state (c, d). Applied shear stress versus time (upper panels) and corresponding viscosity profiles (lower panels) are plotted. e-f) Optical images of SJ state at the time of reversal, 300 s (e) and at 300.6 s (f) showing the contact breaking of MWCNT flocs upon stress reversal. Arrow indicates the direction of applied shear and red circles aid to see contact breaking. Scale bar: 200 $\mu$m.}
		\label{Reversal}
	\end{figure}
	
	We observe shear jamming for low volume fractions $\phi \geq 0.5\%$. Remarkably, we obtain a direct transition to SJ state without any DST regime for the entire range of applied stress and particle volume fractions. To our knowledge, such behaviour has never been observed for suspensions. We will be looking into the details of this atypical feature in the following discussions.  In Fig. \ref{FlowCurve}b, we show the variation of threshold stress for the onset of yielding ($\tau_y$) and shear jamming ($\tau^*$) as functions of volume fraction. The variation of yield stress with volume fraction exhibits a power law scaling, $\tau_{y} \sim \phi^{2.8}$ similar to other flocculated suspensions \cite{buscall1987rheology}. Similarly, the onset stress for SJ, $\tau^* \sim \phi^{2}$. For all the concentrations, we find that the SJ states cannot withstand significant applied stresses. The SJ state typically fails beyond $\sigma \sim$ 15 Pa as also reported earlier \cite{majumdar2011discontinuous}. Such failure at relatively low stress values indicates that the rigidity of the MWCNT flocs is much lower compared to the rigid grains (e.g. corn starch, silica particles).
	\begin{figure}[h]
		\centering
		\includegraphics[width=1\columnwidth]{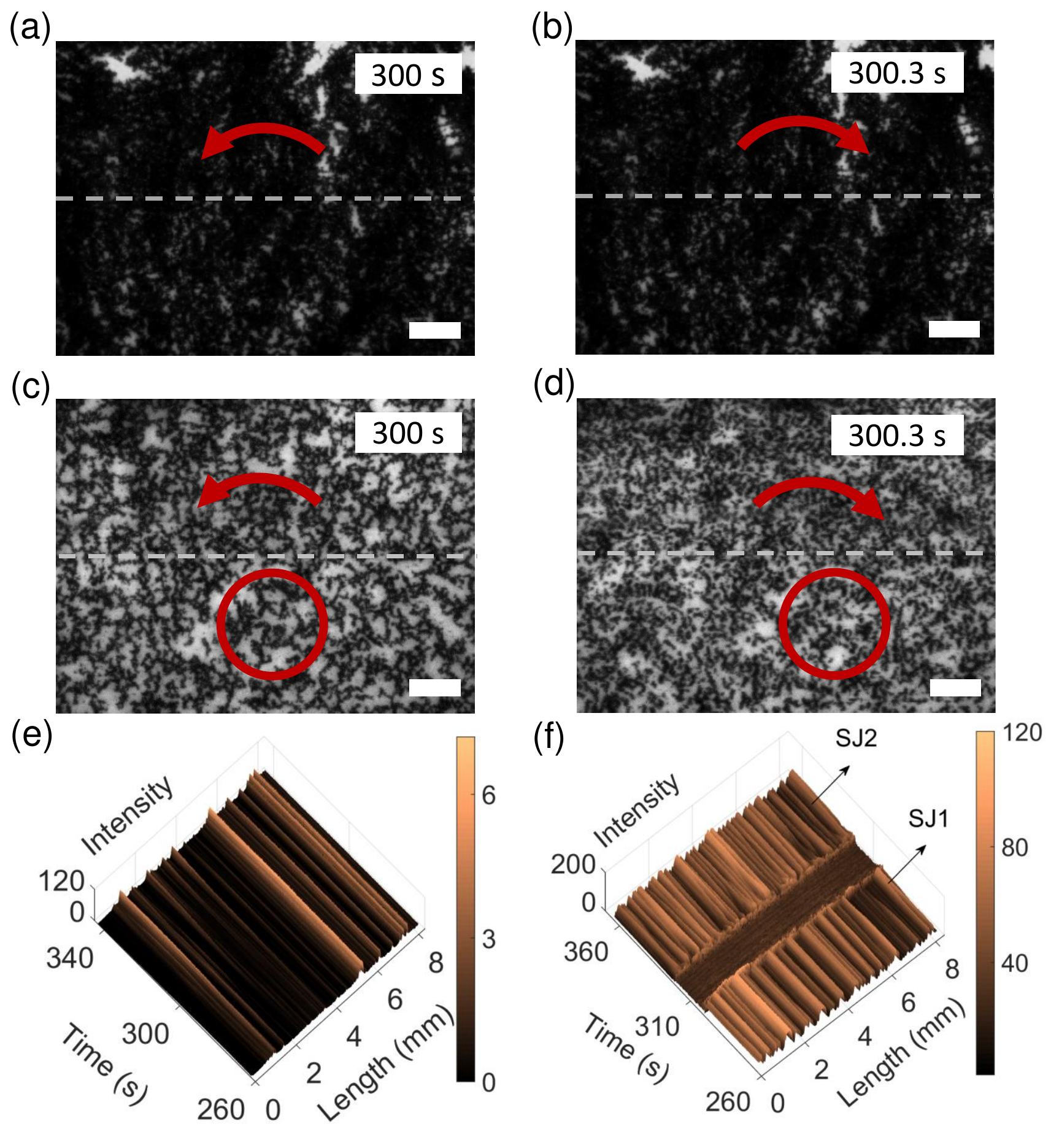}
		\caption{Optical images taken at the time of stress reversal and after 0.3 s for CJ state (a, b) and SJ state (c, d). Scale bar: 1 mm. Arrows indicate the direction of applied shear stress and the red circles aid to see contact breaking upon the stress reversal. e and f) Temporal evolution of the structure of CJ state (e) and SJ state (f), plotted as the variation of light intensity vector at a fixed line (dashed line) on the images.}
		\label{RevImg}
	\end{figure} 
	
	One of the main characteristic features that differentiates a shear jammed state from an isotropically jammed state is the fragility under applied perturbations \cite{cates1998jamming, seto2019shear}. To investigate the fragility and associated structural changes in MWCNT suspensions, we carried out stress reversal experiments coupled with in situ optical imaging. The latter is possible in the present case because shear jamming in MWCNT suspensions starts at much lower volume fractions due to the fractal nature of flocs as compared to the high volume fraction ($\sim$ 55\%) requirement for spherical colloids. This enables us to track the contact dynamics during the SJ transition using in situ optical imaging. We first apply a constant shear stress of 0.02 Pa ($< \tau_y$) for 300 s to the suspension at rest ($\phi$ = 0.77\%). The suspension stays undisturbed in the CJ state. After 300 s, the direction of the stress is reversed as shown in Fig. \ref{Reversal}a. We observe that the viscosity remains unchanged during the reversal (Fig. \ref{Reversal}b) as the suspension remains jammed, confirming the non-fragile, isotropic nature of the CJ state. To investigate the fragility of the SJ state, a constant shear stress of 3 Pa ($>\tau^*$) is applied to a freshly loaded suspension ($\phi$ = 0.77\%) as shown in Fig. \ref{Reversal}c. Initially, the suspension flows for $\sim$ 90 s and enters an SJ state (marked as SJ1) which can support the applied shear stress. The structural transformations associated with this rheological response, captured using in situ optical imaging, are shown in Fig. S4. It shows diffuse MWCNT network structure in the quiescent state, vorticity aligned rolling log-like flocs and dispersed broken flocs at the flowing-state \cite{lin2004elastic, varga2019hydrodynamics} and interconnected dense floc network structure at the SJ state \cite{majumdar2011discontinuous}. After 300 s, the stress direction is reversed which results in an  immediate melting of the SJ1 state characterized by a drop in viscosity by several orders of magnitude (Fig. \ref{Reversal}d). This flowing-state continues for $\sim$ 20 s and then it enters a re-entrant SJ state (marked as SJ2) corresponding to the new direction of shear. Optical images captured before the reversal of stress direction ($t$ = 300 s, SJ1 state) and just after the reversal ($t$ = 300.6 s) are shown in Fig. \ref{Reversal}e and \ref{Reversal}f, respectively. The images clearly show contact breaking between the fractal flocs within a fraction of a second as the SJ state is melted upon the  stress reversal. In contrast, no change in the contact between the flocs is observed in the case of CJ state before and after the stress reversal. Such anisotropy in the SJ state indicates the presence of frictional contacts in the system that can statically counter the applied stress in one direction but becomes unstable on stress reversal \cite{lin2015hydrodynamic, cates1998jamming}. Although the correlation between fragility and contact breaking has been studied for SJ in granular matter \cite{bi2011jamming}, direct experimental observation of such correlation remained unexplored as yet in dense suspensions. In the flowing-state just before shear jamming, the effective volume fraction of the flocs, $\phi_{eff} \sim 36.3 \%$ remains much higher than the actual volume fraction ($\phi = 0.77 \%$) (Fig. S4c). We have estimated $\phi_{eff}$ using optical imaging as described in the S.I.. Such high effective volume ensures that the system can show SJ even at a very low volume fraction of MWCNT. Estimation of the 3-D fractal dimension ($d_f$) of the flocs in the flowing-state, where the individual flocs can be identified, yields a value $d_f \sim$ 2.1 (S.I.).
	\begin{figure}[h]
		\centering
		\includegraphics[width=1\columnwidth]{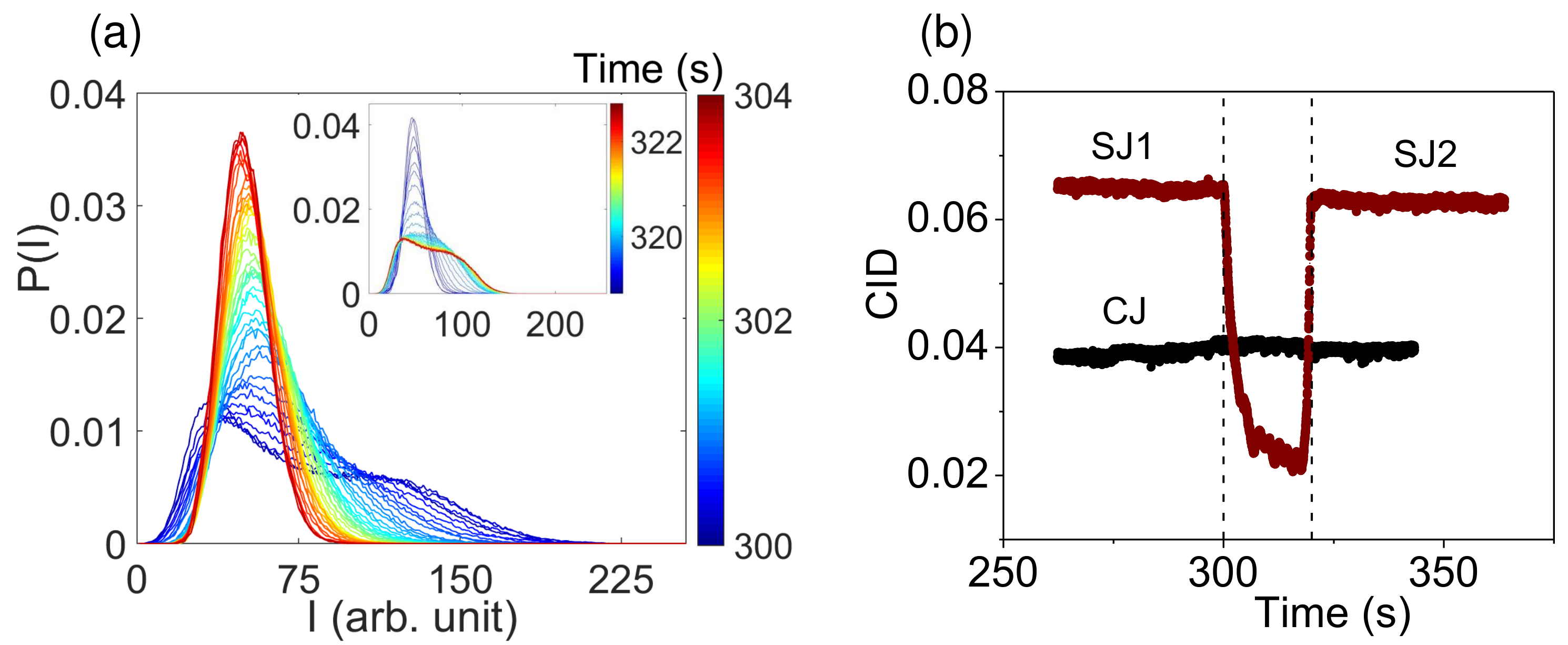}
		\caption{Temporal evolution of probability distribution functions of pixel intensity at the stress reversal (a) and across the re-entrant shear jamming (inset). Color bar represents the time. Inset axis labels are the same as that of the main graph. b) Computable Information Density (CID) variation with time at the stress reversal in CJ state and SJ state. Vertical dash lines indicate the time of reversal (300 s) and that of the re-entrance of SJ state (320 s).}
		\label{PDF}
	\end{figure}
	
	\begin{figure*}[h]
		\centering
		\includegraphics[width=2\columnwidth]{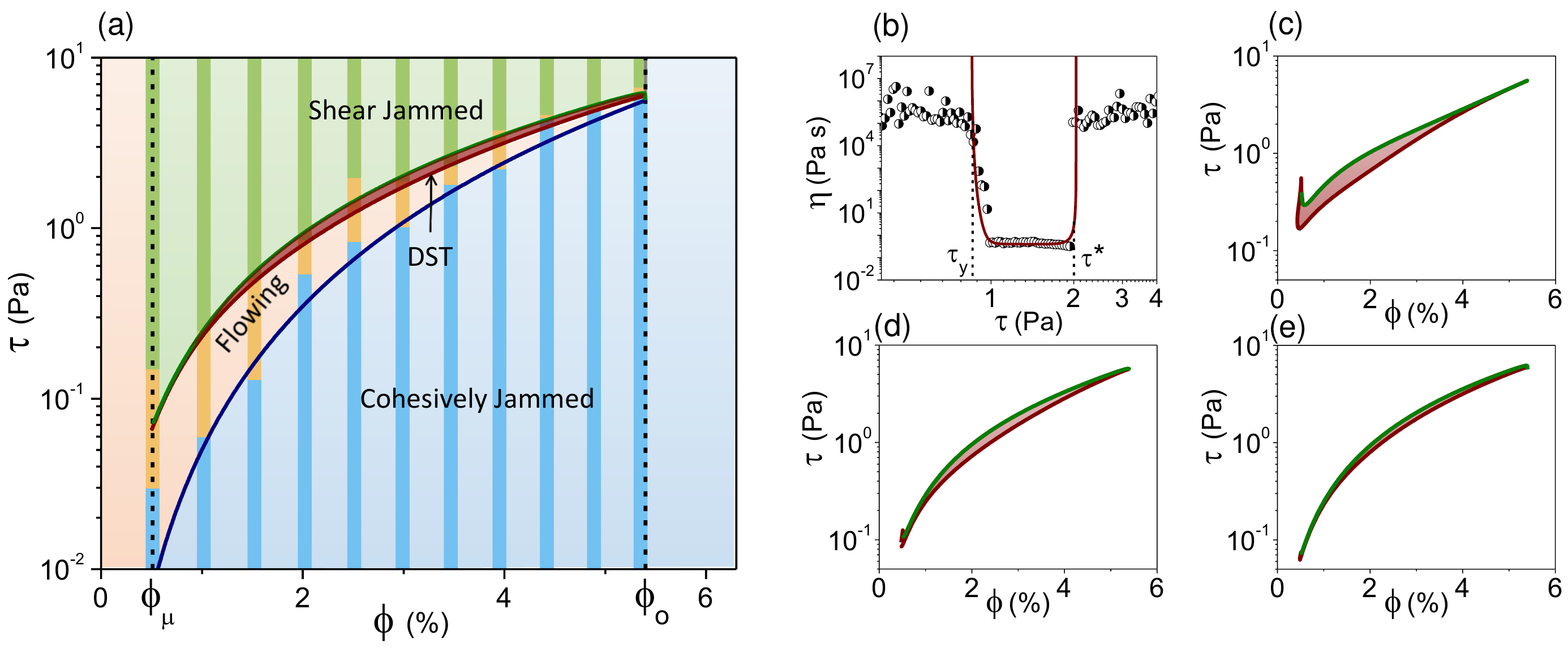}
		\caption{a) The phase diagram for fractal suspensions in $\tau$ - $\phi$ parameter space. Stacked columns represent the experimental data and solid lines denote the phase boundaries extracted from the model for $\alpha = 10$. Different phases marked are cohesively jammed (unyielded), shear thinning/flowing and shear jammed states. The narrow red region indicates the DST phase. b) The flow curve of MWCNT suspension ($\phi$ = 2.5\%) fitted with the combined constitutive model of Herschel-Bulkley and Wyart-Cates. Symbols show the experimental data and the solid line denotes the constitutive model fit. c-e) Narrowing of the DST phase (shaded area) in the phase diagram with respect to growth rate, $\alpha$ = 1, 3 and 10 respectively. Boundaries of DST (red) and shear jamming (green) are shown.}
		\label{PhaseDia}
	\end{figure*}
	
	The isotropic nature of the CJ state and the fragile, anisotropic nature of the SJ state are further confirmed by a large field of view, transmission mode imaging (Fig. \ref{RevImg}). There is no structural change in the CJ state upon stress reversal (Fig. \ref{RevImg}a and \ref{RevImg}b) whereas the floc structure in the SJ state immediately starts to break upon stress reversal (Fig. \ref{RevImg}c and \ref{RevImg}d). Such distinct behaviour of CJ and SJ states can be captured by space-time plots (Fig. \ref{RevImg}e and \ref{RevImg}f) depicting the time evolution of intensity along the dotted lines marked in Figs. \ref{RevImg}a - \ref{RevImg}d. We find that the intensity distribution does not show any time evolution for the CJ state, but for the SJ state a clear change is observed during the transition from the SJ1 to the SJ2 state upon stress reversal (Fig. \ref{RevImg}f). 
	
	High field of view imaging and sampling the dynamics of a large number of flocs enable us to probe the statistics of the stress-induced structural evolution of the SJ state. In Fig. \ref{PDF}a, we show the time evolution of the probability distribution of pixel intensity $P(I)$ recorded in the transmission mode during the stress reversal. Before reversal, $P(I)$ shows a bimodal nature with two distinct peaks at low (the dark region where MWCNT flocs are present) and high (void space) intensities. Immediately after the reversal ($t>$ 300.1), the two peaks start to merge giving rise to a single peak at an intermediate intensity closer to the low intensity peak.  In the transmission mode imaging, the intensity at each pixel of the image is a direct measure of transmitted light intensity through the sample across the gap between the shearing plates: $I = I_0~e^{-\lambda d}$, where $I_0$ is the incident light intensity, $\lambda$ is the absorption coefficient and $d$ is the thickness of MWCNT flocs across the shear gap. Under the assumption that the value of $\lambda$ remains constant for the flocs, the increase in the intensity at any pixel indicates a reduction in the thickness of MWCNT flocs across the gap between shear plates at the corresponding position. Hence, the merging of two intensity bands reflects the dissociation of fractal flocs and redistribution of disjointed flocs in the lateral direction. We observe an exact opposite trend during the recurrence of shear jamming as shown in the inset of Fig. \ref{PDF}a: a single peak of $P(I)$ in the flowing-state splits into two distinct peaks in the re-entrant SJ state, signifying the building up of system spanning contacts.
	
	To further quantify the complex structural transition associated with SJ, we measure the computable information density (CID) that can capture such transitions in a wide range of equilibrium and non-equilibrium systems \cite{martiniani2019quantifying}. We show the variation of CID obtained from the ratio of the compressed ($L'$) to the uncompressed ($L$) image size as a function of time during the stress reversal in Fig. \ref{PDF}b. We observe that in the SJ state CID remains almost constant. Interestingly, CID shows a sharp drop after the stress reversal as the system starts to flow. As the re-entrant SJ state sets in, CID increases again and reaches almost the same value as in the initial SJ state. The strong correlation between the floc structure and SJ is remarkable and novel. We conjecture that the lower value of CID in the flowing-state is due to the uniform distribution of broken flocs as compared to the SJ state which shows larger and distinct fractal structures. No such time-variation is observed under stress reversal for the CJ state as seen in Fig. \ref{PDF}b.
	
	Now we model the rheological behaviour of our system which is strikingly different from the conventional dense suspensions. We require a constitutive relation that incorporates a cohesive jammed (CJ) state for $\tau<\tau_y$, shear thinning for $\tau_y < \tau < \tau^*$ and shear jamming for $\tau > \tau^*$. For our system, yielding from the CJ state to shear thinning state is well described by Herschel-Bulkley (HB) equation, $\tau_{_{\text{HB}}} = \tau_y + k \dot{\gamma}^n$ ($k$: consistency index; $n$: power-law exponent). Since the stress reversal experiments show the existence of frictional contacts in our system, we consider the Wyart-Cates (WC) model \cite{wyart2014discontinuous} which is widely used for capturing DST and SJ in suspensions of frictional particles. In WC model the stress dependence of jamming concentration is given by $\phi_J (\tau) = \phi_\mu f(\tau) + \phi_0 \left(1-f(\tau)\right)$. Here, the function, $f(\tau) \in [0,1]$, indicates the fraction of frictional contacts in the system. At low shear stress, when almost all the contacts are lubricated, $f(\tau) \sim 0$. At higher applied stress, the lubrication layers between the particles are breached to form frictional contacts. When a sufficient number of the contacts are transformed into frictional ones at very high stress, $f(\tau)$ approaches 1. For MWCNT suspensions, the values of $\phi_0$ and $\phi_{\mu}$ are 5.4\% and 0.51\%, respectively. Such stress-induced increase of $f(\tau)$ is empirically modeled by an exponential function \cite{lin2015hydrodynamic, mari2014shear}. The strength of frictional contact at a microscopic scale also affects $f(\tau)$. The sharp rise to the SJ state in our system indicates a rapid increase in the sufficient number of contacts between the flocs.  We describe this behaviour with a stretched exponential function \cite{guy2015towards, dhar2019signature}, $f(\tau) = 1 - e^{-(\frac{\tau-\tau_y}{\tau^*-\tau_y})^\alpha}$ for $\tau > \tau_y$ and $f(\tau) = 0$ for $\tau \leq \tau_y$,	where the exponent $\alpha$ determines the growth rate of $f$ with respect to $\tau$ (see Fig. S5). It ensures that all the contacts are lubricated till the yielding and the frictional contacts start to form at higher stress values. For non-fractal frictional particles, $\alpha \sim 1$. With $\alpha >> 1$, the rapid increase of $f(\tau)$ makes $\phi_J$ to drop towards $\phi_\mu$ giving rise to an abrupt divergence of viscosity. We find that a constitutive model combining Herschel-Bulkley and Wyart-Cates relations for viscosity \cite{singh2019yielding},   
	\begin{equation}
	\label{eqn_comb}
	\eta(\tau, \phi) = \frac{k^{1/n}~ \tau_y}{(\tau - \tau_y)^{1/n}}+\frac{k^{1/n}}{(\tau-\tau_y)^{1/n -1}} + \eta_0 \left(1 - \frac{\phi}{\phi_J}\right)^{-\beta}
	\end{equation}
	where $\eta_0$ is the solvent viscosity, captures the flow curve very well in the limit of $\alpha >> 1$, as shown in Fig. \ref{PhaseDia}b. Unlike earlier models, the onset stress ($\tau^*$) to overcome the lubrication barrier between particles is not a constant for our system but increases with $\phi$ as shown in Fig. \ref{FlowCurve}b. This indicates that the stress-induced frictional contacts are possible only when the applied stress is high enough to overcome the yield stress arising from the cohesive interparticle interactions. The values of parameters $\eta_0$, $n$ and $k$ used for the modeling of experimental data is given in the Supplementary Information.
	
	Based on the combined constitutive model, we map out a phase diagram for fractal suspensions in $\tau-\phi$ parameter space as shown in Fig. \ref{PhaseDia}a. It demarcates different phases, viz. unyielded (CJ state), shear thinning (flowing-state), DST and shear jamming. DST occurs when there is a discontinuous jump in the shear stress at a given shear rate, i.e., $\frac{d\,log(\tau)}{d\,log(\dot{\gamma})} = \infty $ and jamming occurs when $\dot{\gamma} = 0$ at a higher stress value. Yielding and shear jamming boundaries are estimated from Eqn. \ref{eqn_comb}. Interestingly, the growth rate of $f(\tau)$ with shear stress controls the DST and SJ boundaries: the DST regime shrinks as the growth rate of $f(\tau)$ increases. Narrowing of the DST phase in the phase diagram is demonstrated in Fig. \ref{PhaseDia}c-e, for growth rate, $\alpha =$ 1, 3 and 10. With high values of $\alpha$ ($>>$1), the DST regime almost disappears as the DST and SJ boundaries overlap. Consequently, the system can directly transit from flowing to SJ state, bypassing the narrow DST zone. DST is a prominent phase for suspensions of compact particles with well-defined geometries. In contrast, we experimentally observe a direct transition from flowing to SJ phase in MWCNT fractal suspensions at all volume fractions, $\phi_{\mu} \leq \phi \leq \phi_0$. The absence of DST stems from the rapid growth of a sufficient number of interlocking flocs forming a network structure that can easily span the system due to the confinement. Another notable point from the phase diagram is the rise of the shear jamming boundary with $\phi$ showing an opposite trend compared to the conventional, non-fractal systems showing DST and SJ. This is a direct consequence of $\phi$ dependent $\tau_y$ due to cohesive interactions in the system. The power-law exponent in the WC model (Eqn. \ref{eqn_comb}) is taken as $\beta$ = 2, as in the case of conventional systems. We note that the SJ phase boundary is not very sensitive to the value of $\beta$. Although the experimental data show a direct transition from a flowing (yellow region) to SJ state (green region) as indicated in Fig. \ref{PhaseDia}a, we use a moderate value ($\alpha$ = 10) to indicate a narrow DST region. This establishes the generality of the model capturing all the possible phases. However, if we use very high $\alpha$ values ($\alpha\geq$ 50), the DST regime disappears completely.
	
	\begin{figure}
		\includegraphics[width=0.48\textwidth]{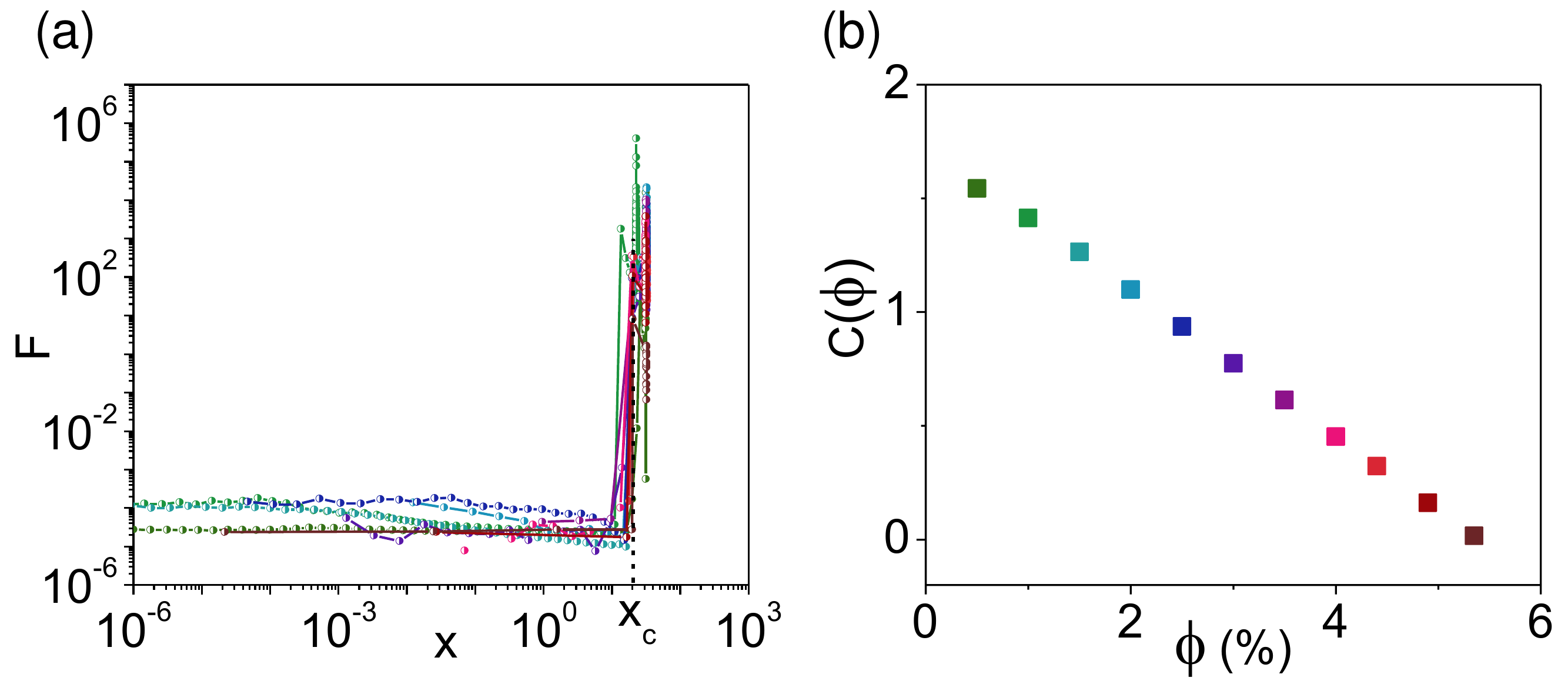}
		\caption{a) Scaling function $F(x) = (\eta-\psi(\phi)) (\phi_0-\phi)^2$ vs. scaling variable $x = \frac{f(\tau) C(\phi)}{(\phi_0-\phi)}$ for 0.5\%$\leq \phi \leq$ 5.35\%. Colors indicate different volume fractions as depicted in panel (b). b) Variation of $C(\phi)$ with volume fraction.}
		\label{UniversalCurve}
	\end{figure}
	
	A very recent study shows that by recasting the WC model, the viscosity of shear thickening suspensions can be collapsed onto a universal curve \cite{ramaswamy2021universal}. The underlying scaling function $F(x)$ which is related to the anisotropy of the system, is obtained from the relation, $(\eta-\psi(\phi)) (\phi_0-\phi)^2 = F(x)$ with the scaling variable, $x = \frac{f(\tau) C(\phi)}{(\phi_0-\phi)}$. Here,  $\eta-\psi(\phi)$ is the reduced viscosity where a volume fraction dependent part of the viscosity, $\psi(\phi)$, ensures the data collapse below the shear-thickening/jamming onset. Variation of $F(x)$ vs $x$ for a range of $\phi$ values is shown in Fig. \ref{UniversalCurve}a, where we obtain a good scaling of the data. The scaling function $F(x)$ diverges at $x = x_c = (\phi_0 - \phi_{\mu})^{-1}$. In our case, $x_c \approx 20.4$. We find from Fig. \ref{UniversalCurve}b that $C(\phi)$ decreases linearly with increasing $\phi$ values and eventually vanishes at the isotropic jamming point, $\phi_0$. This linear decrease in $C(\phi)$ is similar to that observed for the shear-thickening suspensions of silica particles \cite{ramaswamy2021universal}. As mentioned in ref. \cite{ramaswamy2021universal}, the function $C(\phi)$ is related to the volume fraction dependent anisotropy of the SJ state.
	
	\section{Conclusion}
	We have presented stress-induced divergence of viscosity in fractal MWCNT suspensions showing all the signatures of shear jamming at very low concentrations ($\phi_\mu \sim$ 0.5\%). By combining stress reversal experiments along with in situ optical imaging, we bring out the fragility of the SJ state and associated contact dynamics and structural reorganization. Such correlations have remained unexplored so far in the experimental studies of shear jamming in dense suspensions. Interestingly, from in situ optical imaging, we observe that the effective area coverage of fractal clusters is significantly reduced in the SJ state as compared to the initial CJ state (see Fig. \ref{RevImg}). This indicates that at stress, $\tau > \tau_y$, the flow induces densification of the flocs making them stiffer which facilitates the contact network formation required to achieve the SJ state. For dense suspensions of frictional non-fractal particles, when the phase diagram is plotted in the parameter plane of stress ($\tau$) and particle volume fraction ($\phi$) \cite{peters2016direct}, the DST regime narrows down as $\phi \rightarrow \phi_0$. However, in our case, we observe a direct transition from flowing to SJ with no CST/DST regime for all volume fractions ($\phi$: 0.5 \% - 5.4 \%) up to the isotropic jamming point ($\phi_0$ = 5.4 \%), as shown in Fig. 5a. This signifies that the SJ phenomenon reported here is quite different from that of compact granular systems. The very high effective friction related to the stress-induced interlocking of fractal clusters is responsible for such striking stress response. 
	
	Importantly, we present a phase diagram for fractal suspensions based on the generalized Wyart-Cates model showing a cohesive jammed state, shear thinned state and shear jammed state. The generality of the phase diagram sets a framework for the rheology of fractal suspensions. The sharp increase in the number of frictional contacts over a narrow range of stress is the key reason to observe a shear jammed state from the flowing state without going through the DST phase as commonly observed in dense suspensions of spherical or anisotropic particles. Controlling fractal dimension in these systems by tuning the inter-particle interactions and observing the resulting flow behaviour can shed light on the sharp rise in stress-induced contact formation or contact proliferation. Exploring the stability properties of SJ state in fractal suspensions using a similar approach used for dry granular materials \cite{zhao2022ultrastable} can be an interesting future direction to explore.  We hope that our study will motivate further experimental and theoretical studies of SJ in suspensions of fractal particles. 
	
	\section*{Acknowledgements}
	AKS thanks the Department of Science and Technology, India for the support under Year of Science Professorship and the Nanomission Council. SM thanks Science and Engineering Research Board, India for the support through Ramanujan Fellowship.
	
	\section*{Materials and Methods}
	For the preparation of suspensions, MWCNTs having a diameter of 30-50 nm and length of a few microns purchased from Sun NanoTech are mixed with an organic solvent, N-methyl-2-pyrrolidone (NMP), from Sigma Aldrich. Suspensions of different volume fractions, $\phi = 0.26 - 5.35\%$, are prepared by a continuous process of mechanical stirring (10 min) and ultra-sonication (10 min) followed by mechanical stirring again (10 min). This preparation protocol is implemented to ensure a homogeneous distribution of flocs in the medium.  The volume fraction of MWCNT suspension is calculated from the  weight fraction using density of NMP (1.03 g/cm\textsuperscript{3}) and the true density of MWCNT (1.7 - 2.1 g/cm\textsuperscript{3}).
	
	Rheological measurements are carried out in a stress-controlled rheometer (MCR 102, Anton Paar) at a fixed temperature of 25 $^{o}$C. Flow curves are obtained using a sandblasted cone-plate measuring system (diameter: 50 mm, cone-angle: 2$^{o}$) and a steel bottom plate. Opto-rheological measurements are done on suspensions ($\phi$ = 0.77$\%$) using a glass parallel plate measuring system (diameter: 43 mm) along with a glass bottom plate at a fixed shear gap of 60 $\mu$m. Images are captured using a color CCD camera (Luminera, 640 X 480 pixels) fitted with a microscope objective lens. A white light source is used for illuminating the imaging area. Images are collected at a rate of 50 fps in transmission mode (Leica lens 5x) and in reflection mode (Mitutoyo lens 5x with condenser). Optical images of drop cast samples are taken using Olympus bright-field microscope.
	
	
	\balance
	
	
	\providecommand*{\mcitethebibliography}{\thebibliography}
\csname @ifundefined\endcsname{endmcitethebibliography}
{\let\endmcitethebibliography\endthebibliography}{}

	\fontsize{11}{12}\selectfont
	\onecolumn
	\doublespacing
	\setcounter{figure}{0}
	\makeatletter 
	\renewcommand{\thefigure}{S\@arabic\c@figure}
	\makeatother
	\begin{center}
		\LARGE \bf{{Supplementary Information}}
	\end{center}
	\vspace{0.25cm}
	
	\section*{Fractal structure of MWCNT flocs}
\begin{figure}[h]
		\begin{center}
			\includegraphics[width=0.6\textwidth]{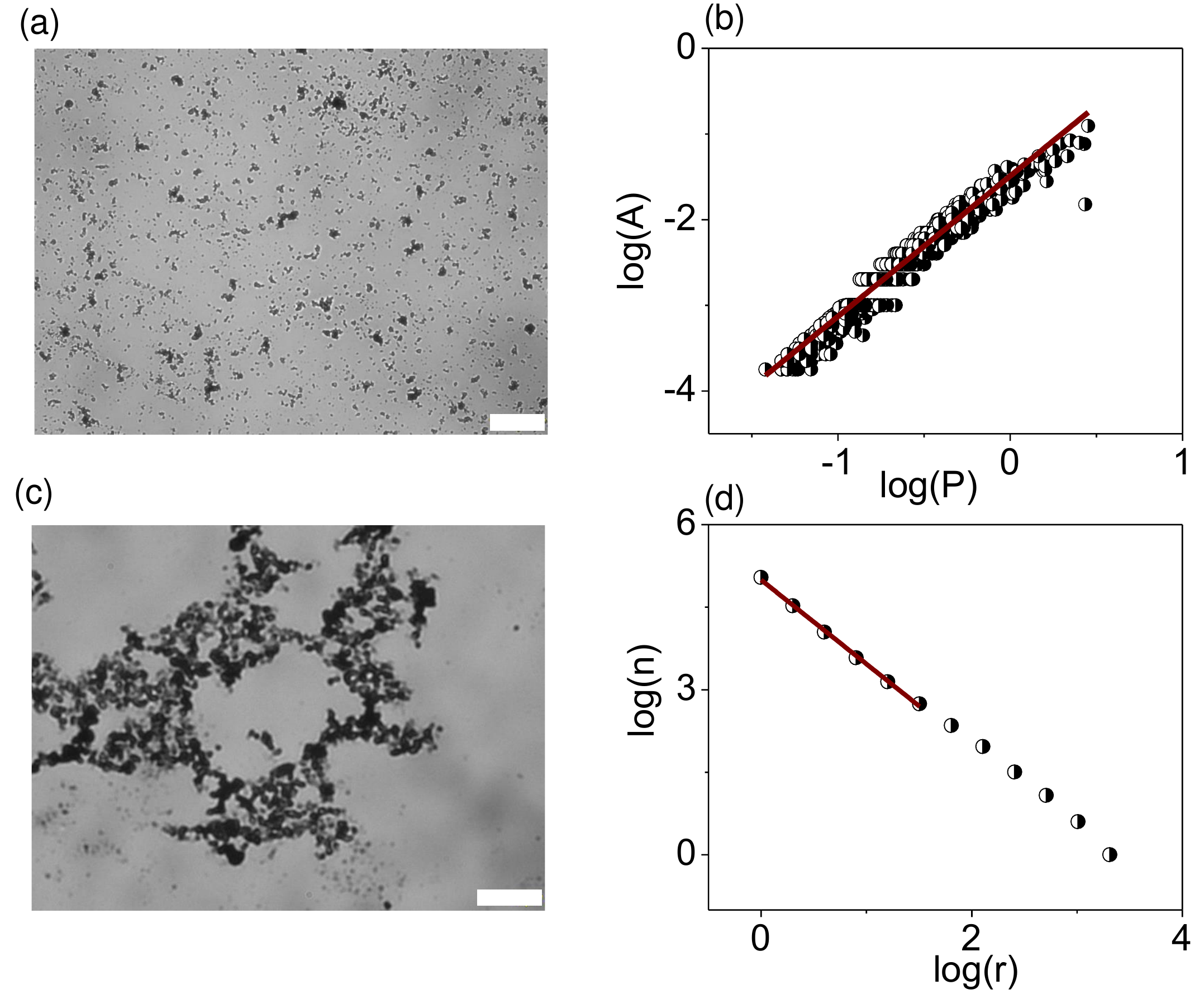}
			\caption{a) Optical image of discrete MWCNT fractal flocs (scale bar: 50 $\mu$m). b) Log(A) versus log(P) of the individual flocs and the linear fit. c) Optical image of the interconnected fractal flocs (Scale bar: 20 $\mu$m). d) Log(n) versus log(r) of the interconnected flocs and the linear fit.}
			\label{Fractal}
		\end{center}
	\end{figure}

	For the fractal structure analysis, MWCNT suspension ($\phi$ = 0.03$\%$) is drop cast onto a glass substrate and dried. Optical microscopic images of the drop cast samples show the fractal nature of MWCNT flocs. The 2D fractal dimension ($d_f$) is estimated by two methods: Area-Perimeter analysis ($log(A) \propto d_f~log(P)$) and box-counting method ($d_f = -\lim_{r \to 0} \frac{log(n)}{log(r)}$  where $n$ is the number of filled boxes and $r$ is the box size). The Area-Perimeter analysis using the image of discrete flocs (Fig. \ref{Fractal}a) yields the value $d_f = 1.64$ (see Fig. \ref{Fractal}b). For the box-counting method, images of interconnected flocs are collected in which the fractal arrangement of nanotubes is clear. A typical image and the fractal dimension estimation ($d_f = 1.53$) are shown in Fig. \ref{Fractal}c and d respectively.  As we increase the concentration for drop casting, such analysis becomes difficult due to the dense packing of the dried clusters. So, we have used a very dilute sample ($\phi$ = 0.03$\%$) for such analysis.
	
	The concentration of the drop cast sample is low compared to that of the sample used for rheology measurements. At those higher concentrations, it becomes difficult to image the fractal arrangement of nanotubes due to clustering. To address this issue, we have directly estimated the 3D fractal dimension of the suspension used for the rheology measurement from the effective volume fraction: $\phi_{eff} \approx \phi (R/a)^{3 - d_f}$, where $R$ is the radius of gyration and $a$ is the size of MWCNT \cite{Genovese}. Using the image of dispersed broken flocs in the flowing state (Fig. S4c), the estimated value of 3D fractal dimension is, $d_f \approx 2.1$ ($\phi_{eff} = 36.3\%$, $\phi = 0.77\%$, $R = 100\mu$m and $a = 1\mu$m). Here, we have considered only the flowing state of the sample just beyond yielding, where the isolated clusters are observed that enables us to estimate the value of $R$. In the CJ and SJ states, the determination of $R$ is difficult using optical imaging due to the connectivity of the clusters. For the estimation of effective volume fraction, the area coverage of the dispersed broken flocs (Fig. S4c) is measured using ImageJ software. Assuming a uniform distribution of flocs across the shear gap, $\phi_{eff}$ is estimated.	
	
	\section*{Flow Curves of MWCNT suspensions}
\begin{figure}
		\begin{center}
			\includegraphics[width=1\textwidth]{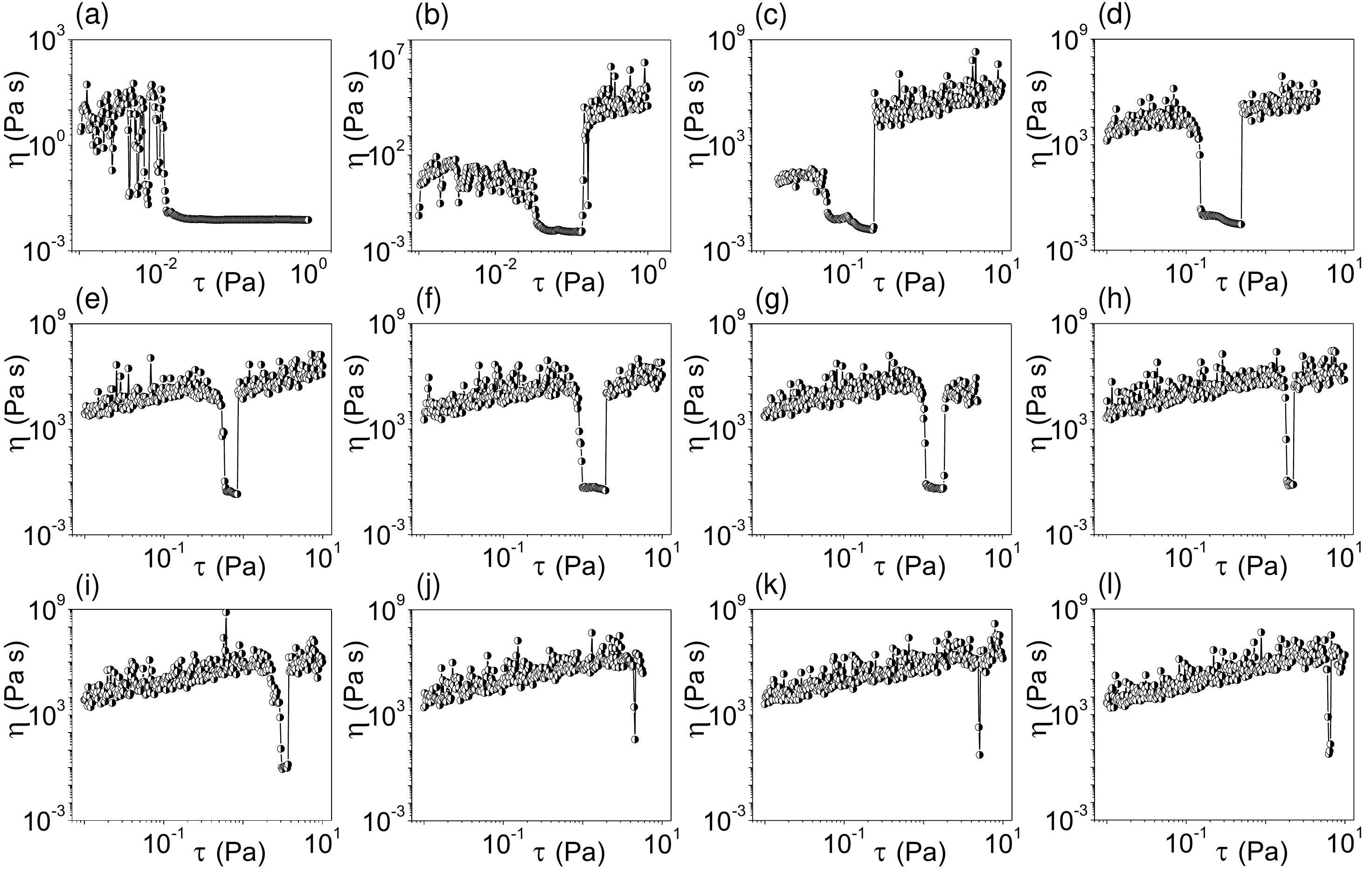}
			\caption{a - l) Flow curves of MWCNT suspensions of different volume fractions, 0.26, 0.51, 1.02, 1.52, 2.01, 2.51, 2.99, 3.47, 3.95, 4.42, 4.89, 5.35$\%$ respectively, plotted as viscosity versus shear stress in log-log scale.}
			\label{AllFC}
		\end{center}
	\end{figure}

	Flow curves of MWCNT suspensions of different volume fractions ranging from 0.26$\%$ to 5.35$\%$ are plotted as viscosity versus applied shear stress in Fig. \ref{AllFC}. Shear jamming transition is observed for $\phi \geq 0.5 \%$. From the variation of viscosity with shear stress, $\eta \sim \tau^{\delta}$, the onset of DST is characterized with, $\delta = 1$. Here, the divergence of viscosity at finite shear stress and the corresponding drop in shear rate, $\dot{\gamma} \sim 0$, imply direct manifestation of shear jamming without prior DST.
	
\begin{figure}
		\begin{center}
			\includegraphics[width=0.7\textwidth]{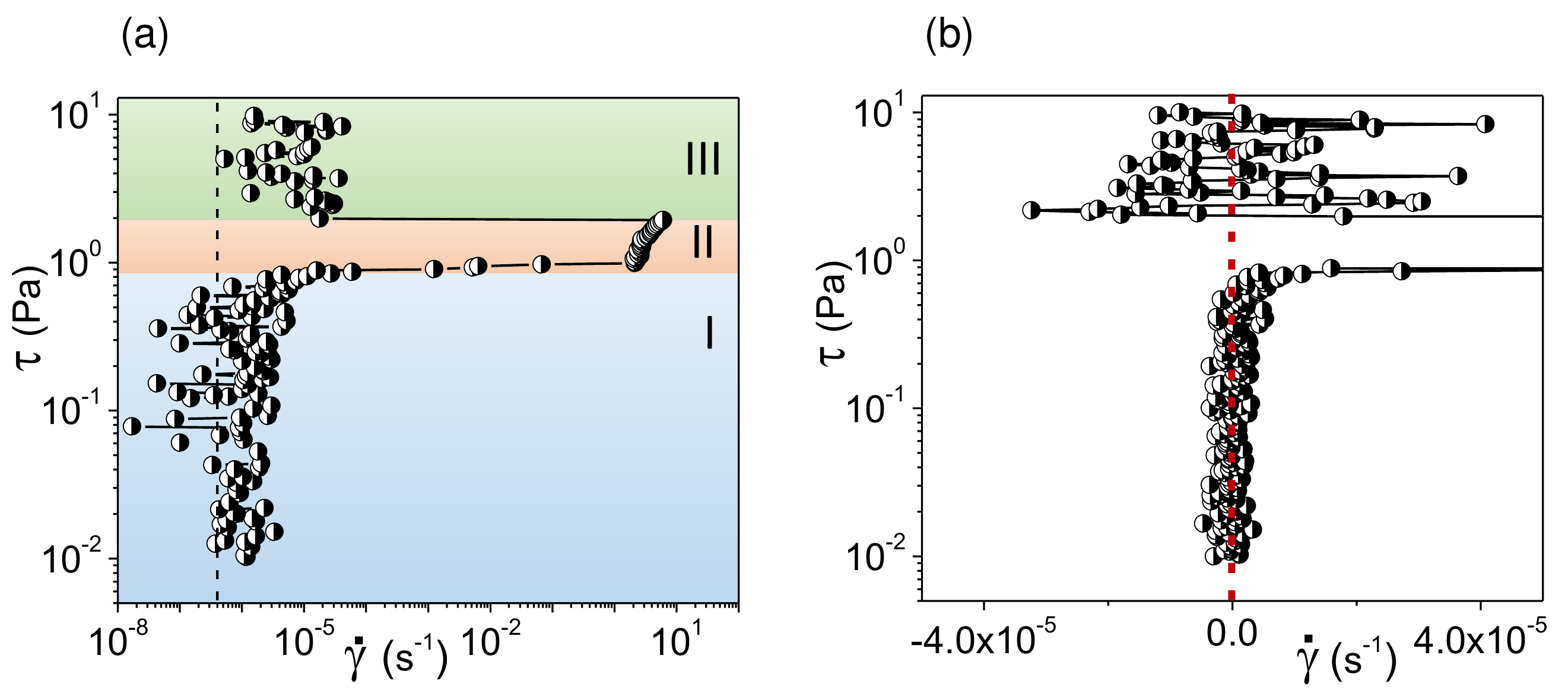}
			\caption{a) S-shaped flow curve of MWCNT suspension ($\phi$ = 2.5\%), plotted as shear stress versus shear rate in log-log scale. The dashed line denotes the shear rate at the resolution limit of the rheometer. b) Magnified portion of the flow curve showing the fluctuation of shear rate around zero (vertical dashed line) in both CJ and SJ states.}
			\label{Zoom}
		\end{center}
	\end{figure}

	Since MWCNT suspensions belong to the category of frictional suspensions which exhibit DST or SJ or both, we use S-shaped flow curve representation \cite{Wyart} of the rheological data in Fig. \ref{Zoom}a. In the S-shaped flow curve, SJ is indicated by arch-shapes with a vanishing $\dot{\gamma}$ at higher stress. A magnified portion of the flow curve ($\phi$ = 2.5\%) is shown in Fig. \ref{Zoom}b. It clearly shows that the shear rate drops down to the resolution limit of the rheometer and fluctuates around zero in regions I and III. Since the average shear rate is negligible in these regions, they are solid-like jammed states; namely, cohesive jammed state and shear jammed state.
	
	\section*{Stress reversal images}
\begin{figure}
		\begin{center}
			\includegraphics[width=1\textwidth]{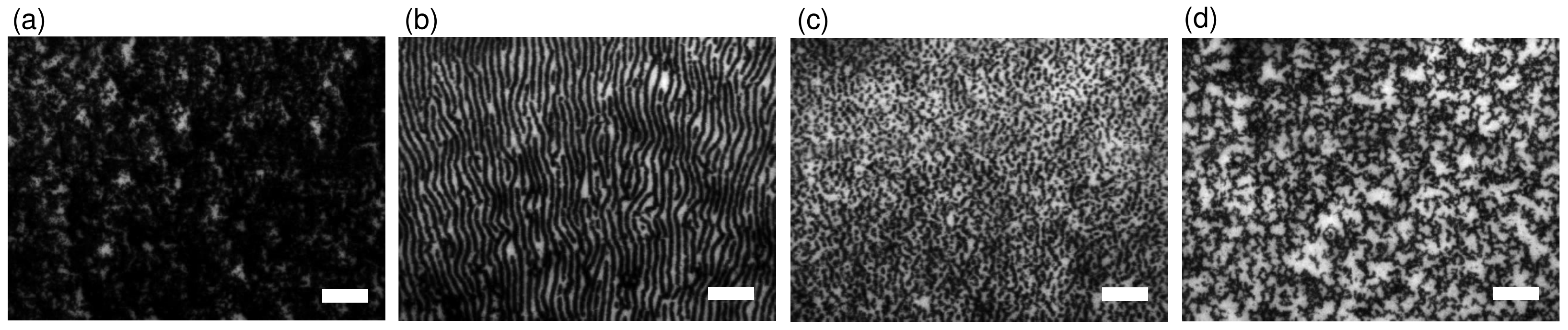}
			\caption{Optical images taken after the application of shear stress (3 Pa). It shows diffuse MWCNT network structure at the quiescent state (a), rolling log-like flocs (b) and dispersed broken flocs (c) at the flowing-state and interconnected dense floc network structure at the SJ state (d). Scale bar: 1 mm.}
			\label{Structure}
		\end{center}
	\end{figure}

	In situ optical images of MWCNT suspension ($\phi = 0.77\%$) collected in transmission mode during the stress reversal with a wide-angle lens (Leica, 5x) are shown in Fig. \ref{Structure}. It shows the structural transformations associated with the initial flow of suspensions when shear stress of 3 Pa is applied. The structural deformation pathways leading to the shear jamming can be clearly seen in the images. 
	
	\section*{Parameters for the modeling}
	For modeling the rheology data of MWCNT suspensions using the combined constitutive model of Herschel-Bulkley and Wyart-Cates, we used parameter values, $n = 0.2$, $k = 0.1$ and $\eta_0 = 1.65$ mPa s. The critical volume fraction of jamming without friction, $\phi_0$ = 5.4\%, is the volume fraction above which the system remains in the CJ state under the application of shear stress. The critical volume fraction of jamming with friction, $\phi_{\mu}$ = 0.51\%, is the volume fraction above which the system exhibits shear jamming.
	
	\section*{$f(\tau)$ growth at different rates}
\begin{figure}
		\begin{center}
			\includegraphics[width=0.4\textwidth]{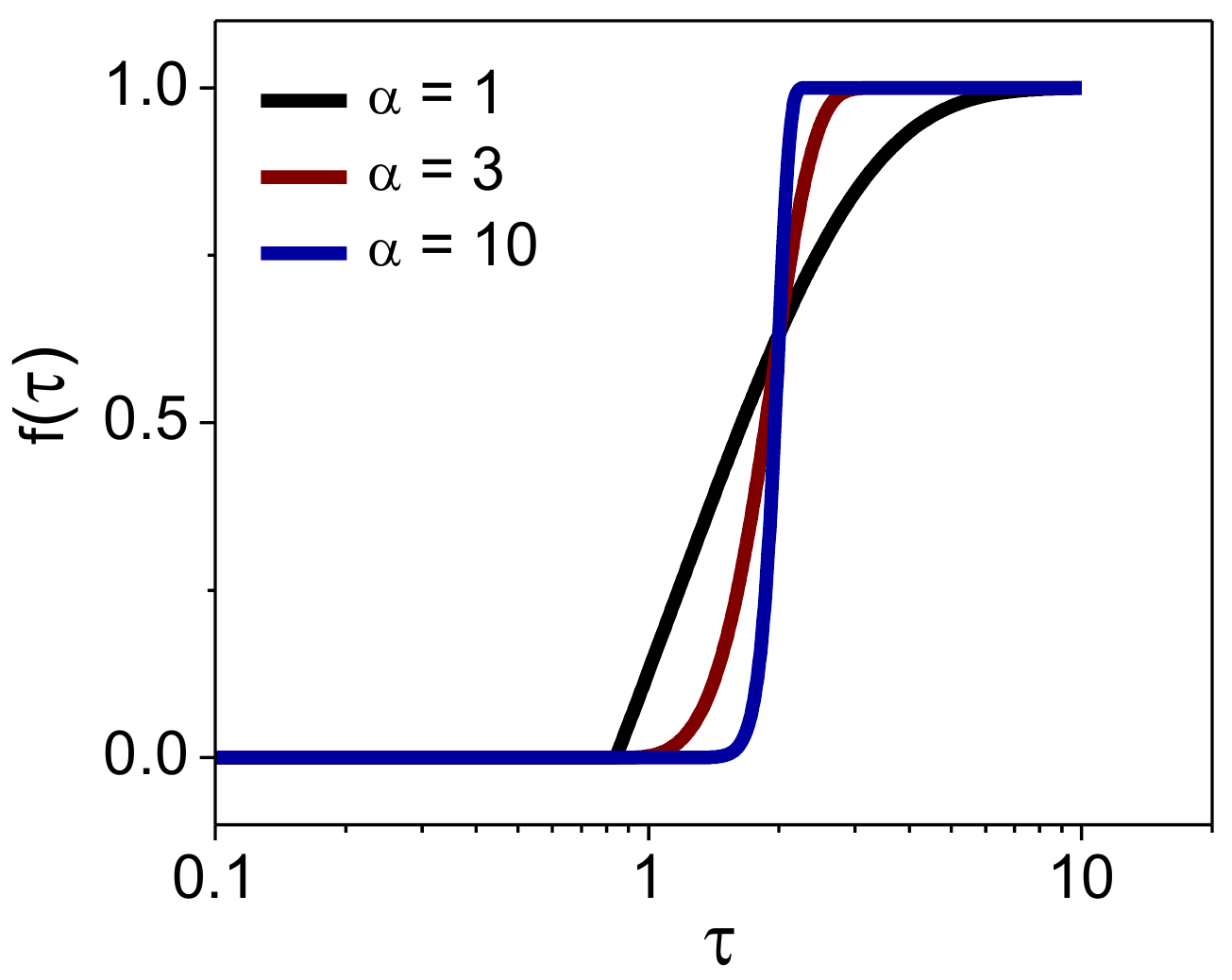}
			\caption{Growth of $f(\tau)$ with shear stress for different growth rates, $\alpha$ = 1, 3 and 10.}
			\label{Foftau}
		\end{center}
	\end{figure}

	Growth of fraction of frictional contacts with shear stress, $f(\tau) = 1 - e^{-(\frac{\tau-\tau_y}{\tau^*-\tau_y})^\alpha}$ for $\tau > \tau_y$ and $f(\tau) = 0$ for $\tau \leq \tau_y$, for different values of growth rates, $\alpha$ = 1, 3 and 10, is shown in Fig. \ref{Foftau}.

\end{document}